%% file: feyn.tex
\documentclass{pzstyle}
\usepackage{pzcmd}
\hypersetup{%
 pdftitle={jQuery.Feyn: Drawing Feynman Diagrams with SVG},
 pdfauthor={Zan Pan},
 pdfsubject={Software},
 pdfkeywords={Feynman diagrams, jQuery, SVG}
}

\begin{document}

\title{\textbf{jQuery.Feyn: Drawing Feynman Diagrams with SVG}}
\author{Zan Pan\thanks{\href{mailto:panzan@itp.ac.cn}{panzan@itp.ac.cn}}
 \\[4pt] Institute of Theoretical Physics,\\
 Chinese Academy of Sciences, Beijing, 100190, P.R.~China}
\date{\today}
\maketitle

\begin{abstract}
 jQuery.Feyn is a tool for drawing Feynman diagrams with Scalable Vector Graphics (SVG),
 written in JavaScript and runs in modern browsers. It features predefined
 propagator styles, vertex types, and symbols. Math formulae can be included as
 external graphics, or typeset with TeX through MathJax library. The generated
 SVG code can be easily modified to make fine adjustments and conveniently
 transferred using copy-and-paste.
\end{abstract}

\input{intro.tex}

\input{usage.tex}

\input{design.tex}

\input{summary.tex}

\appendix
\input{options.tex}

\input{bezier.tex}

\input{references.tex}

\end{document}

%% file: intro.tex
\section{Introduction}

In the field of high energy physics, Feynman diagrams are widely used as
pictorial representations of the interaction of sub-atomic particles\cite{DK05}.
However, it is not easy to draw such a complicated object in publications
(Interestingly, Feynman slash notation is also not easy to typeset).
This has led to the development of computer programs. In the framework of LaTeX,
there are four approaches: Michael Levine's \texttt{feynman}\cite{ML90} bundle,
Jos Vermaseren's \texttt{axodraw}\cite{JV94} package, Thorsten Ohl's
\texttt{feynmf}\cite{TO95} package, and Norman Gray's \texttt{feyn}\cite{NG09}
font. Besides, Binosi \emph{et. al.}'s \texttt{JaxoDraw}\cite{BT04, BCKT08}
and Hahn and Lang's \texttt{FeynEdit}\cite{HL08} are both written in Java
which provide graphical user interfaces.

The reason why we designed another program is that publishing with
HTML5\cite{SK11} has grown to be a very promising technology and requires a tool
running in browsers to produce visually pleasing Feynman diagrams for the Web
such as Wikipedia or online scientific articles. jQuery.Feyn is an atempt to
fill that gap. Compared with the programs mentioned above, our program has
great advantages in flexibility and portability.

\subsection{Overview}

jQuery.Feyn is a \href{http://jquery.com/}{jQuery} plugin to facilitate
drawing Feynman diagrams with SVG\cite{JD02}. It makes full use of
jQuery's succinctness and extensibility to embrace the tagline:
\emph{write less, do more}. The following provides a summary of
jQuery.Feyn's main features:
\begin{itemize}
 \item Automatic generation of clean SVG source code
 \item Easy to use, easy to make fine adjustments
 \item Predefined propagator styles, vertex types, and symbols
 \item Support for typesetting labels and including external graphics
 \item Lightweight, cross-browser, and fully documented
\end{itemize}

The home of jQuery.Feyn project is \url{http://photino.github.io/jquery-feyn/}.
Please refer to this link for up-to-date documentation and practical examples.

\subsection{Supported Browsers}

Any modern browsers for desktop or mobile with a basic support of inline SVG in HTML5
in the standards mode should be OK to run jQuery.Feyn. However, we do not guarantee
that all of them will display the same SVG exactly on your screen due to their
disparities in SVG rendering and support level. Also note that mobile browsers
often show a lot of quirks that are hard to work around because of their
limitations and different UI assumptions. The following provides an incomplete
list of supported browsers:
\begin{itemize}
 \item Firefox 4+
 \item Chrome 7+
 \item Opera 11.6+
 \item Safari 5.1+
 \item IE 9+
\end{itemize}

Personaly, I recommend Firefox 24+ and Chrome 28+, on which my testing will be
conducted continually. There is no doubt that newer browsers always have
better support for SVG and hence better support for jQuery.Feyn.

\subsection{Bug Reports and Comments}

The preferred way to report bugs is to use the GitHub issue tracker:
\begin{center}
 \url{https://github.com/photino/jquery-feyn/issues}
\end{center}
Of course, you can also email me at \href{mailto:panzan@itp.ac.cn}{panzan@itp.ac.cn}
or \href{mailto:panzan89@gmail.com}{panzan89@gmail.com}. When reporting bugs,
you should be as specific as possible about the problem so that we can easily
reproduce it. If possible, please test them on Firefox 24+ and Chrome 28+
to get rid of your browser's quirks. Comments, questions, and requests for
adding more features are also welcome.

\subsection{License}

\begin{verbatim}
Copyright (C) 2013 by Zan Pan <panzan89@gmail.com>

Permission is hereby granted, free of charge, to any person obtaining a copy
of this software and associated documentation files (the "Software"), to deal
in the Software without restriction, including without limitation the rights
to use, copy, modify, merge, publish, distribute, sublicense, and/or sell
copies of the Software, and to permit persons to whom the Software is
furnished to do so, subject to the following conditions:

The above copyright notice and this permission notice shall be included in
all copies or substantial portions of the Software.

THE SOFTWARE IS PROVIDED "AS IS", WITHOUT WARRANTY OF ANY KIND, EXPRESS OR
IMPLIED, INCLUDING BUT NOT LIMITED TO THE WARRANTIES OF MERCHANTABILITY,
FITNESS FOR A PARTICULAR PURPOSE AND NONINFRINGEMENT. IN NO EVENT SHALL THE
AUTHORS OR COPYRIGHT HOLDERS BE LIABLE FOR ANY CLAIM, DAMAGES OR OTHER
LIABILITY, WHETHER IN AN ACTION OF CONTRACT, TORT OR OTHERWISE, ARISING FROM,
OUT OF OR IN CONNECTION WITH THE SOFTWARE OR THE USE OR OTHER DEALINGS IN
THE SOFTWARE.
\end{verbatim}

%% file: usage.tex
\section{Usage}

\subsection{Getting Started}

You can start your tour from jQuery.Feyn's online demo:
\begin{center}
 \url{http://photino.github.io/jquery-feyn/demo.html}
\end{center}
Settings for Feynman diagrams are in the form of JavaScript's liberal object notation,
whose simplicity and extensibility has already made it possible for jQuery.Feyn
to be both lightweight and powerful. we hope you will enjoy this feature if it is
still unfamiliar to you. Knowledge on jQuery is not necessary to use jQuery.Feyn,
but getting a feel of JavaScript syntax will make your exploration easier. More importantly,
you should be familiar with SVG markup language\cite{JD02}. Unlike PostScript,
the SVG code has great readability. It is not a problem to grasp it in a short time.

To use jQuery.Feyn, the first thing you should do is to load the scripts
found in the distribution:
\begin{Verbatim}
<script src="js/jquery-2.0.2.min.js"></script>
<script src="js/jquery.feyn-1.0.0.min.js"></script>
\end{Verbatim}
Please note that jQuery library comes first. After this, you can proceed to
configure your desired Feynman diagram like
\begin{Verbatim}[frame=single,rulecolor=\color{brown},%
 xleftmargin=3mm,numbers=left,numbersep=4pt]
<script>
  $(document).ready(function() {
    $("#container").feyn({
      incoming: {i1: "20,180", i2: "180,180"},
      outgoing: {o1: "20,20", o2: "180,20"},
      vertex: {v1: "100,140", v2: "100,60"},
      fermion: {line: "i1-v1-i2,o2-v2-o1"},
      photon: {line: "v1-v2"}
    });
  });
</script>
\end{Verbatim}

\begin{figure}[!ht]
 \centering
 \ifpdf
  \includegraphics[scale=0.8]{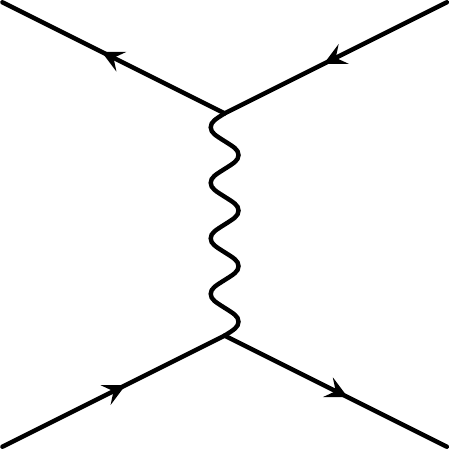}
 \else
  \includegraphics[scale=0.8]{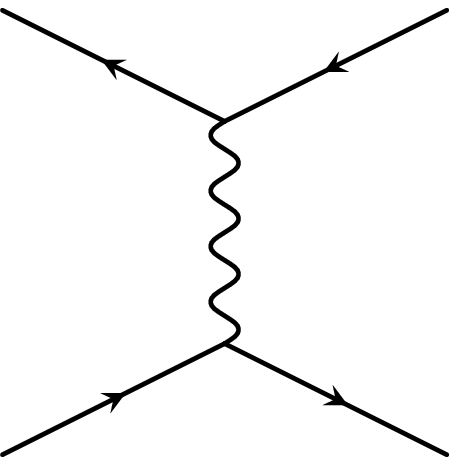}
 \fi
 \caption{The minimal example of jQuery.Feyn's output\label{fig:minimal}}
\end{figure}

The jQuery ID selector \verb|$("#container")| can also be replaced by any
other selector that selects a unique block-level element in the document,
which serves as the container of jQuery.Feyn's SVG output. The minimal example
illustrated above represents a QED process (see Figure~\ref{fig:minimal}).

As can be seen, jQuery.Feyn has done the most part of work automatically.
If you are unsatisfied with the output of jQuery.Feyn, or if you would like to
add a graphics element that are not provided by jQuery.Feyn, you can always
modify the SVG code manually. By setting the \texttt{standalone} option
to \texttt{true}, you can edit the source code for your diagram in the textarea.

\subsection{Default Options}

The default options of jQuery.Feyn's \texttt{Feyn} constructor
are listed as follows

\begin{Verbatim}
{
  xmlns: "http://www.w3.org/2000/svg",
  xlink: "http://www.w3.org/1999/xlink",
  version: "1.1",
  x: 0,
  y: 0,
  width: 200,
  height: 200,
  title: "",
  description: "Feynman diagram generated by jQuery.Feyn",
  standalone: false,
  selector: false,
  grid: {show: false, unit: 20},
  color: "black",
  thickness: 1.6,
  tension: 1,
  ratio: 1,
  clockwise: false,
  incoming: {},
  outgoing: {},
  vertex: {},
  auxiliary: {},
  fermion: {arrow: true},
  photon: {period: 5, amplitude: 5},
  scalar: {arrow: false, dash: "5 5", offset: 2},
  ghost: {arrow: true, thickness: 3, dotsep: 8, offset: 5},
  gluon: {width: 15, height: 15, factor: 0.75, percent: 0.6, scale: 1.15},
  symbol: {},
  node: {show: false, thickness: 1, type: "dot", radius: 3, fill: "white"},
  label: {family: "Georgia", size: 15, face: "italic"},
  image: {},
  mathjax: false,
  ajax: false
}
\end{Verbatim}

As you have alreaday seen, they can be overridden by passing an option object
to \texttt{feyn} method. However, options are not checked in any way,
so setting bogus option values may lead to odd errors. For JavaScript
internal errors excluding syntax errors, jQuery.Feyn will write the error
information to the container of your Feynman diagram to remind you of the case.
A complete list of available options will be given in the
appendix~\ref{sec:options}.

\subsection{Tips, Tricks, and Troubleshooting}

\begin{itemize}
 \item You can edit the generated SVG code directly to make fine adjustments.
  The corresponding SVG output will be updated immediately when you
  trigger a text change event by clicking outside of the textarea.
  It is your own responsibility to ensure the validity of your SVG code.
  Note that reloading the page will descard your editing, so please manually
  save the change in an external SVG file. You can use the two textareas
  in this way: one for testing, and the other for producing.
 \item Simple labels such as particle names, momentum, data, and comments, can be
  typeset with jQuery.Feyn's \texttt{label} option. It supports subscript,
  superscript, bar and tilde accents by using the \texttt{dx} and \texttt{dy}
  attributes of \texttt{<tspan>}. For complicated mathematical expressions,
  they should be included as external SVG images with the \texttt{image} option.
  Troy Henderson's \href{http://www.tlhiv.org/ltxpreview/}{LaTeX Previewer}
  provides a user-friendly utility for generating LaTeX output\cite{TH12}.
 \item For special characters such as greek letters and mathematical operators,
  it is recommended to input the unicode entity by citing its decimal number,
  for example $\alpha$ can be accessed by \verb|&#945;|. A list of frequently
  used characters can be found at
  \href{http://www.ascii-code.com/html-symbol.php}{ascii-code.com/html-symbol}.
 \item If you are familiar with Mathematical Markup Language (MathML), you can
  also include mathematical expressions by adding the \texttt{<foreignObject>}
  element manually. Please check \href{http://caniuse.com/mathml}{caniuse.com/mathml}
  to see whether or not your browser has a good support for MathML.
 \item When \href{http://www.mathjax.org/}{MathJax} is available, you can
  set jQuery.Feyn's \texttt{mathjax} option to \texttt{true} to typeset
  mathematics in TeX or LaTeX. This functionality also relies on browsers'
  support for the \texttt{<foreignObject>} element.
 \item SVG files can be converted to
  \href{http://image.online-convert.com/convert-to-eps}{EPS} and
  \href{http://document.online-convert.com/convert-to-pdf}{PDF} online.
  If your SVG code has included some external SVG files, please set
  jQuery.Feyn's \texttt{ajax} option to \texttt{true} to merge their
  code directly, or copy, paste, and modify them manually.
  Before conversion, you should \href{http://validator.w3.org/}{check}
  the markup validity of your SVG code.
 \item Chrome does not support loading local files with ajax by default.
  You should start Google Chrome with the \texttt{--disable-web-security}
  or \texttt{--allow-file-access-from-files} option, otherwise you will
  get a network error.
 \item Firefox 23 or below has a
  \href{https://bugzilla.mozilla.org/show_bug.cgi?id=600207}{bug} of
  renderging the \texttt{<image>} element. Please update your browser to 24+.
\end{itemize}

%% file: design.tex
\section{Design}

In this section, we will discuss the design (and implementation) of jQuery.Feyn
from a user's perspective. Before drawing a Feynman diagram, you are encouraged
to draft it on graph paper first.

\begin{figure}[!ht]
 \centering
 \ifpdf
  \includegraphics[scale=0.8]{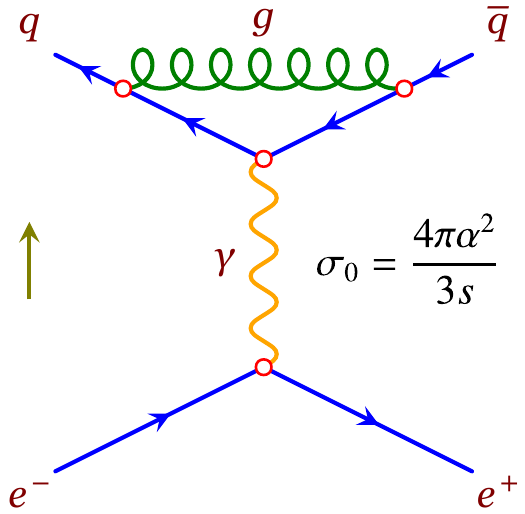} \hspace{20pt}
  \includegraphics[scale=0.8]{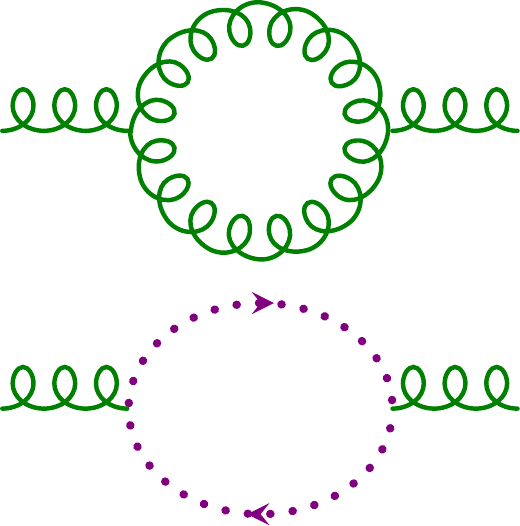} \hspace{20pt}
  \includegraphics[scale=0.8]{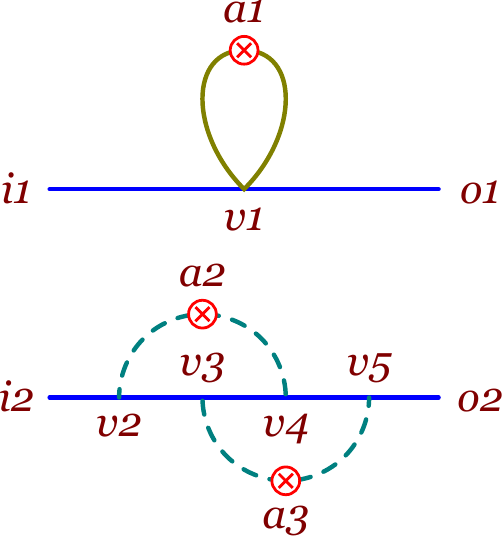}
 \else
  \includegraphics[scale=0.8]{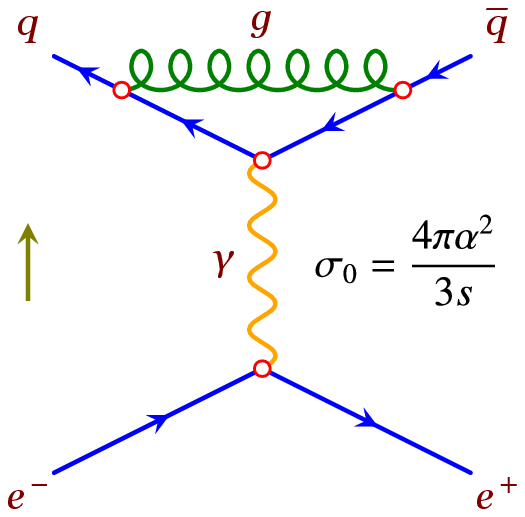} \hspace{20pt}
  \includegraphics[scale=0.8]{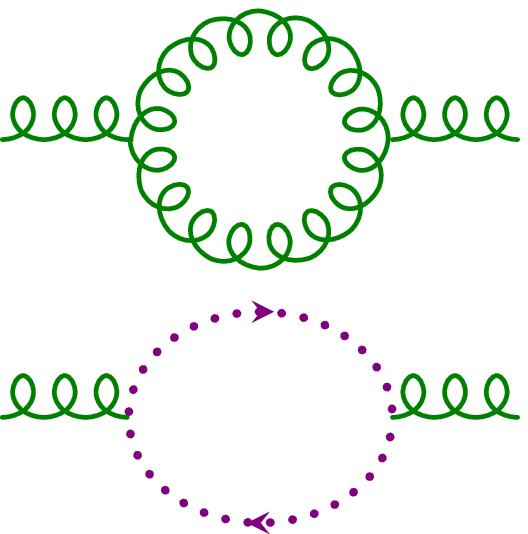} \hspace{20pt}
  \includegraphics[scale=0.8]{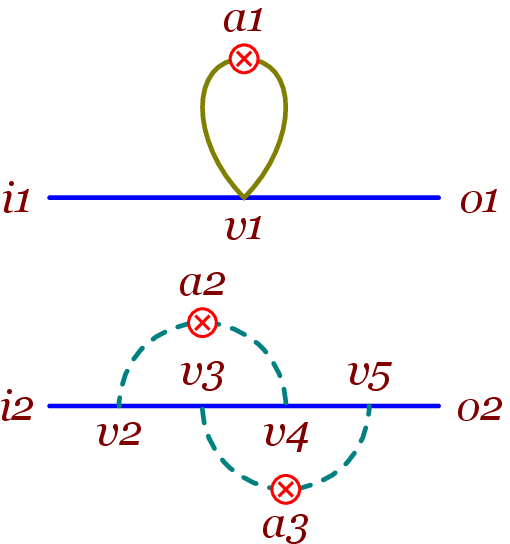}
 \fi
 \caption{Illustrations on jQuery.Feyn's building blocks\label{fig:blocks}}
\end{figure}

Roughly speaking, a Feynman diagram can be treated as a directed or
undirected graph consisting of a set of nodes and edges. As illustrated
in Figure~\ref{fig:blocks}, jQuery.Feyn has provided
five types of basic building blocks that can be used to construct
Feynman diagrams. Each type may contain one or more graphics
primitives (\emph{i.e.} the options):
\begin{itemize}
 \item \textbf{Graph node}: sets the coordinates of nodes with
  the \texttt{incoming}, \texttt{outgoing}, \texttt{vertices} and
  \texttt{auxiliary} primitives, or draws the node marks with
  the \texttt{node} primitive
 \item \textbf{Propagator}: draws the propagators with the \texttt{fermion},
  \texttt{photon}, \texttt{scalar}, \texttt{ghost}, and \texttt{gluon} primitives
 \item \textbf{Symbol}: draws symbols such as arrows, blobs, and so on with
  the \texttt{symbol} primitive
 \item \textbf{Label}: typesets labels with the \texttt{label} primitive
 \item \textbf{Image}: includes external graphics with the \texttt{image} primitive
\end{itemize}

\begin{figure}[!ht]
 \centering
 \ifpdf
  \includegraphics[scale=0.6]{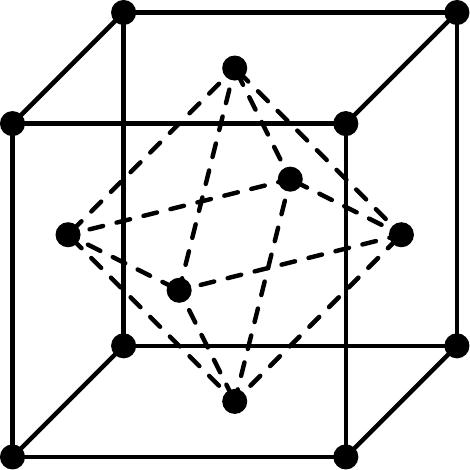} \hspace{20pt}
  \includegraphics[scale=0.6]{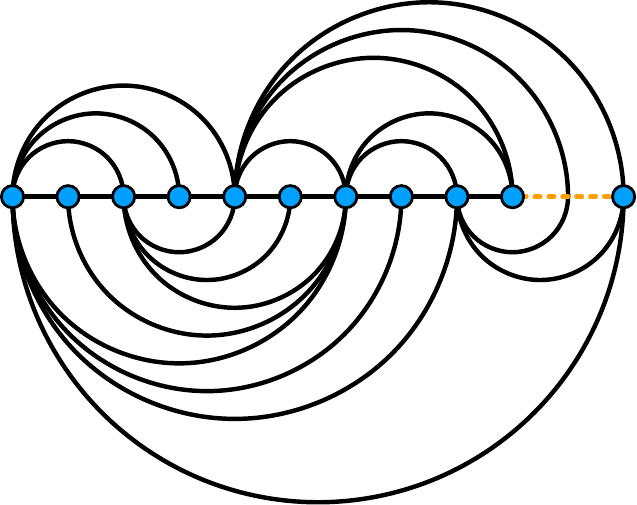}
 \else
  \includegraphics[scale=0.6]{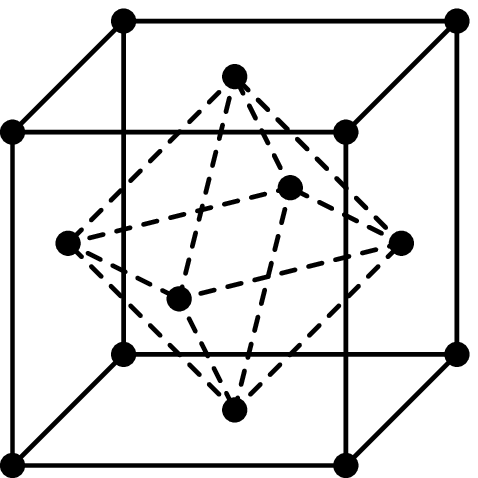} \hspace{20pt}
  \includegraphics[scale=0.6]{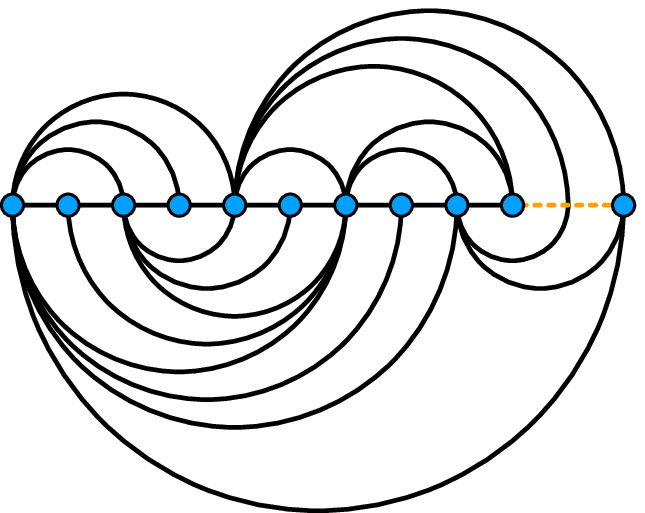}
 \fi
 \caption{The cube with an octahedron inside and the Goldner-Harary graph
  drawn by jQuery.Feyn\label{fig:misc}}
\end{figure}

For the design of line styles and symbols, we also have references to
PSTricks\cite{TZ03}, MetaPost\cite{JH13},
and Asymptote\cite{BHP04}. Beyond Feynman diagrams, jQuery.Feyn also excels
at drawing some mathematical diagrams with dense connections between nodes
(see Figure~\ref{fig:misc}).

\subsection{Graph Nodes}
A graph node is a coordinate pair extracted from the position string such as
\texttt{"20,180"} with a name with the prefix \texttt{i}, \texttt{o},
\texttt{v}, or \texttt{a}. They are respectively the first-letter of
\texttt{incoming}, \texttt{outgoing}, \texttt{vertices}, and \texttt{auxiliary}.
These terminologies should be self-explanatory and are not discussed further.
In fact, all graphics nodes will be merged into one object in the base implementation.
We retain this separation in the user interface just for semantic clarity.

\begin{figure}[!ht]
 \centering
 \ifpdf
  \includegraphics[scale=1.0]{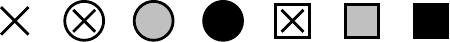}
 \else
  \includegraphics[scale=1.0]{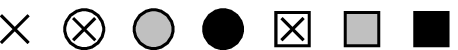}
 \fi
 \caption{Examples of node marks provided by jQuery.Feyn\label{fig:nodes}}
\end{figure}

To draw node marks, you should use the \texttt{node} primitive. Five types of
node marks are provided. Of course, you can also specify diffrent filled colors
to denote different vertices (see Figure~\ref{fig:nodes}).

\subsection{Propagators}
We support five types of propagators: \texttt{fermion}, \texttt{photon},
\texttt{scalar}, \texttt{ghost}, and \texttt{gluon}. According to the conventions
by Peskin and Schroeder\cite{PS95}, only fermion and ghost propagators
show arrows by default. In practice, boson propagators are represented by
the sine curves and gluon propagators by elliptical arcs. They can be
approximated by cubic B\'{e}zier paths in SVG, which will be
discussed in the appendix~\ref{sec:bezier}.

\begin{figure}[!ht]
 \centering
 \ifpdf
  \includegraphics[scale=1.2]{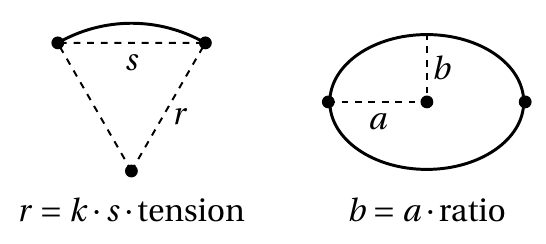}
 \else
  \includegraphics[scale=1.2]{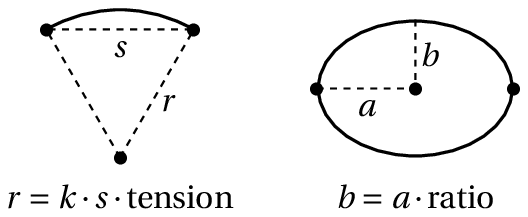}
 \fi
 \caption{The geometrical layout of the arc and loop propagator
  ($k$ is a constant)\label{fig:geometry}}
\end{figure}

Each type of the propagators has three shapes: \texttt{line}, \texttt{arc},
and \texttt{loop}. The \texttt{tension} parameter controls the shape of
arc radius for arc propagators; whereas the \texttt{ratio} parameter controls
the shape of elliptical arc for fermion, scalar, and ghost loop propagators,
\emph{i.e.} the ratio of y-radius to x-radius. Their geometrical meanings
are shown in Figure~\ref{fig:geometry}.

\subsection{Symbols}

\begin{figure}[!ht]
 \centering
 \ifpdf
  \includegraphics[scale=0.8]{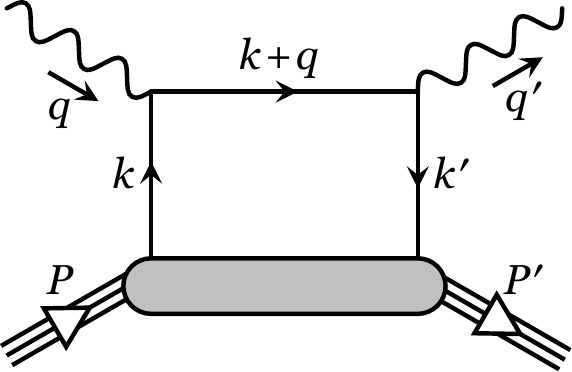}
 \else
  \includegraphics[scale=0.8]{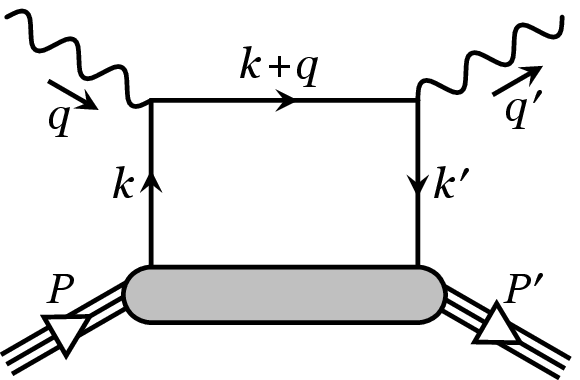}
 \fi
 \caption{Feynman diagram for deeply virtual Compton scattering drawn
  by jQuery.Feyn\label{fig:dvcs}}
\end{figure}

For the convenience of drawing complex diagrams, a small set of predefined
symbols are provided: \texttt{arrow}, \texttt{blob}, \texttt{bubble},
\texttt{condensate}, \texttt{hadron}, and \texttt{zigzag}. Some of them
can also support variants. Examples are illustrated in Figure~\ref{fig:blocks}
and Figure~\ref{fig:dvcs}.

\subsection{Labels}

\begin{figure}[!ht]
 \centering
 \ifpdf
  \includegraphics[scale=0.8]{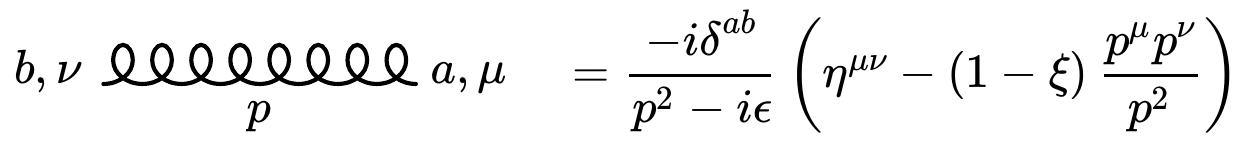}
 \else
  \includegraphics[scale=0.8]{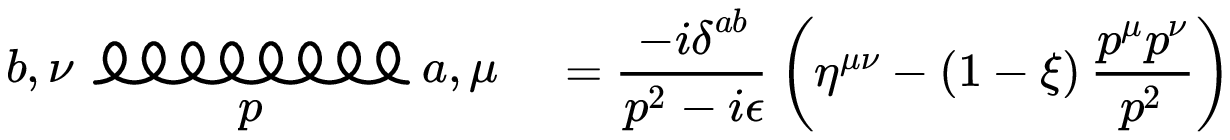}
 \fi
 \caption{Feynman diagram for the gluon propogator drawn
  by jQuery.Feyn with MathJax support\label{fig:gluon}}
\end{figure}

Labels are somewhat annoying. The \texttt{label} primitive has a nice support
for subscript, superscript, and accents. For simple text or annotations such as
particle names and momentum, this is enough. But it is not competent to typeset
math formulae. Two solutions are available: to include as external graphics
(see the first diagram in Figure~\ref{fig:blocks}), or to use the MathJax library
(see Figure~\ref{fig:gluon}). The first method makes use of the \texttt{image}
primitive and always works, while the other depends on browsers' support for
the \texttt{foreignObject} element.

\subsection{Images}

The \texttt{image} primitive is provided to enhance jQuery.Feyn's extensibility.
You can use it to include any external graphics whatever you like. As seen in
Figure~\ref{fig:blocks}, we have embeded the math formula as an SVG file
in the first diagram. It is your own duty to ensure that proper width and
height values are assigned to the image object.

%% file: summary.tex
\section{Summary}

jQuery.Feyn is a tool to draw Feynman diagrams in browsers, which utilizes
the power of SVG. We have explained how to use it and presented some examples
of its building blocks. It is always hard to reconcile ease-of-use with
expressiveness. We encourage the users to modify the generated code manually
to make fine adjustments. As an open standard with the history over a decade,
SVG has rich graphic features and effects, which makes it full of fun to learn
in depth. This also contributes to the reason why we introduce jQuery.Feyn
to physicists for preparing their publications.

\section*{Acknowledgements}

We are grateful to GitHub to host our project. We also acknowledge all the people
who send us feedback during the testing phase.

%% file: options.tex
\section{Complete Reference of Options\label{sec:options}}
In the following, we provide a complete list of jQuery.Feyn's options.
\texttt{String}, \texttt{Number}, \texttt{Boolean}, \texttt{Array},
and \texttt{Object} are JavaScript's data types. Also note that
the bold text in the parentheses is the corresponding default value.

\begin{description}
 \item[xmlns] : \texttt{String} ( \textbf{"http://www.w3.org/2000/svg"} ) \\
  Sets the \texttt{xmlns} attribute of \texttt{<svg>} which binds the SVG namespace
 \item[xlink] : \texttt{String} ( \textbf{"http://www.w3.org/1999/xlink"} ) \\
  Sets the \texttt{xlink:href} attribute of \texttt{<svg>} which defines an IRI reference type as a URI
 \item[version] : \texttt{String} ( \textbf{"1.1"} ) \\
  Sets the \texttt{version} attribute of \texttt{<svg>} which indicates the SVG language version
 \item[x] : \texttt{Number} ( length | \textbf{0} ) \\
  Sets the \texttt{x} attribute of \texttt{<svg>} which indicates a x-axis coordinate in the user coordinate system
 \item[y] : \texttt{Number} ( length | \textbf{0} ) \\
  Sets the \texttt{y} attribute of \texttt{<svg>} which indicates a y-axis coordinate in the user coordinate system
 \item[width] : \texttt{Number} ( length | \textbf{200} ) \\
  Sets the \texttt{width} attribute of \texttt{<svg>} which indicates a horizontal length in the user coordinate system
 \item[height] : \texttt{Number} ( length | \textbf{200} ) \\
  Sets the \texttt{height} attribute of \texttt{<svg>} which indicates a vertical length in the user coordinate system
 \item[title] : \texttt{String} ( text ) \\
  Sets the content for the \texttt{<title>} element which displays a title for the Feynman diagram
 \item[description] : \texttt{String} ( text | \textbf{"Feynman diagram generated by jQuery.Feyn"} ) \\
  Sets the content for the \texttt{<desc>} element which describes the Feynman diagram
 \item[standalone] : \texttt{Boolean} ( true | \textbf{false} ) \\
  Enables or disables the SVG code editor to make fine adjustments and save as a standalone SVG file
 \item[selector] : \texttt{Boolean} ( true | \textbf{false} ) \\
  Determines whether or not to set \texttt{id} and \texttt{class} attributes for SVG elements
 \item[grid] : \texttt{Object}
  \begin{description}
   \item[show] : \texttt{Boolean} ( true | \textbf{false} ) \\
     Determines whether or not to display a grid system to facilitate your drawing
   \item[unit] : \texttt{Number} ( length | \textbf{20} ) \\
    Sets the length of subdivision for the grid system
  \end{description}
 \item[color] : \texttt{String} ( paint | \textbf{black} ) \\
  Sets global \texttt{stroke} attribute for SVG elements which defines the color of the outline
 \item[thickness] : \texttt{Number} ( length | \textbf{1.6} ) \\
  Sets global \texttt{stroke-width} attribute for SVG elements which specifies the width of the outline
 \item[tension] : \texttt{Number} ( parameter | \textbf{1} ) \\
  Sets global parameter of arc radius and the zigzag amplitude
 \item[ratio] : \texttt{Number} ( parameter | \textbf{1} ) \\
  Sets global parameter of elliptical arcs for fermion, scalar, and ghost loop propagators
 \item[clockwise] : \texttt{Boolean} ( true | \textbf{false} ) \\
  Sets global \texttt{clockwise} parameter for propagators
 \item[incoming] : \texttt{Object}
  \begin{description}
   \item[i1, i2, i3, \ldots] : \texttt{String} ( position ) \\
    Sets the coordinate pairs of graph nodes for incoming particles
  \end{description}
 \item[outgoing] : \texttt{Object}
  \begin{description}
   \item[o1, o2, o3, \ldots] : \texttt{String} ( position ) \\
    Sets the coordinate pairs of graph nodes for outgoing particles
  \end{description}
 \item[vertex] : \texttt{Object}
  \begin{description}
   \item[v1, v2, v3, \ldots] : \texttt{String} ( position ) \\
    Sets the coordinate pairs of graph nodes for vertices
  \end{description}
 \item[auxiliary] : \texttt{Object}
  \begin{description}
   \item[a1, a2, a3, \ldots] : \texttt{String} ( position ) \\
    Sets the coordinate pairs of graph nodes for miscellaneous symbols
  \end{description}
 \item[fermion] : \texttt{Object}
  \begin{description}
   \item[color] : \texttt{String} ( paint | \textbf{inherient} ) \\
    Sets the \texttt{stroke} attribute for \texttt{<g>} into which fermion propagators are grouped
   \item[thickness] : \texttt{Number} ( length | \textbf{inherient} ) \\
    Sets the \texttt{stroke-width} attribute for \texttt{<g>} into which fermion propagators are grouped
   \item[tension] : \texttt{Number} ( parameter | \textbf{inherient} ) \\
    Sets the parameter of arc radius for fermion propagators
   \item[ratio] : \texttt{Number} ( parameter | \textbf{inherient} ) \\
    Sets the parameter of elliptical arcs for fermion propagators
   \item[arrow] : \texttt{Boolean} ( \textbf{true} | false ) \\
    Determines whether or not to show arrows for fermion propagators
   \item[clockwise] : \texttt{Boolean} ( true | \textbf{false} ) \\
    Sets the direction of arrows for arc and loop fermion propagators
   \item[line] : \texttt{String} ( connections ) \\
    Sets the directed edges between graph nodes for fermion lines
   \item[arc] : \texttt{String} ( connections ) \\
    Sets the directed edges between graph nodes for fermion arcs
   \item[loop] : \texttt{String} ( connections ) \\
    Sets the directed edges between graph nodes for fermion loops
  \end{description}
 \item[photon] : \texttt{Object}
  \begin{description}
   \item[color] : \texttt{String} ( paint | \textbf{inherient} ) \\
    Sets the \texttt{stroke} attribute for \texttt{<g>} into which photon propagators are grouped
   \item[thickness] : \texttt{Number} ( length | \textbf{inherient} ) \\
    Sets the \texttt{stroke-width} attribute for \texttt{<g>} into which photon propagators are grouped
   \item[tension] : \texttt{Number} ( parameter | \textbf{inherient} ) \\
    Sets the parameter of arc radius for photon propagators
   \item[clockwise] : \texttt{Boolean} ( true | \textbf{false} ) \\
    Determines whether the first wiggle starts up or down for photon propagators
   \item[period] : \texttt{Number} ( parameter | \textbf{5} ) \\
    Sets the period parameter for photon propagators
   \item[amplitude] : \texttt{Number} ( parameter | \textbf{5} ) \\
    Sets the amplitude parameter for photon propagators
   \item[line] : \texttt{String} ( connections ) \\
    Sets the directed edges between graph nodes for photon lines
   \item[arc] : \texttt{String} ( connections ) \\
    Sets the directed edges between graph nodes for photon arcs
   \item[loop] : \texttt{String} ( connections ) \\
    Sets the directed edges between graph nodes for photon loops
  \end{description}
 \item[scalar] : \texttt{Object}
  \begin{description}
   \item[color] : \texttt{String} ( paint | \textbf{inherient} ) \\
    Sets the \texttt{stroke} attribute for \texttt{<g>} into which scalar propagators are grouped
   \item[thickness] : \texttt{Number} ( length | \textbf{inherient} ) \\
    Sets the \texttt{stroke-width} attribute for \texttt{<g>} into which scalar propagators are grouped
   \item[tension] : \texttt{Number} ( parameter | \textbf{inherient} ) \\
    Sets the parameter of arc radius for scalar propagators
   \item[ratio] : \texttt{Number} ( parameter | \textbf{inherient} ) \\
    Sets the parameter of elliptical arcs for scalar propagators
   \item[arrow] : \texttt{Boolean} ( true | \textbf{false} ) \\
    Determines whether or not to show arrows for scalar propagators
   \item[clockwise] : \texttt{Boolean} ( true | \textbf{false} ) \\
    Sets the direction of arrows for arc and loop scalar propagators
   \item[dash] : \texttt{String} ( dasharray | \textbf{"5 5"} ) \\
    Sets the \texttt{stroke-dasharray} attribute for \texttt{<g>} into which scalar propagators are grouped
   \item[offset] : \texttt{Number} ( length | \textbf{2} ) \\
    Sets the \texttt{stroke-offset} attribute for \texttt{<g>} into which scalar propagators are grouped
   \item[line] : \texttt{String} ( connections ) \\
    Sets the directed edges between graph nodes for scalar lines
   \item[arc] : \texttt{String} ( connections ) \\
    Sets the directed edges between graph nodes for scalar arcs
   \item[loop] : \texttt{String} ( connections ) \\
    Sets the directed edges between graph nodes for scalar loops
  \end{description}
 \item[ghost] : \texttt{Object}
  \begin{description}
   \item[color] : \texttt{String} ( paint | \textbf{inherient} ) \\
    Sets the \texttt{stroke} attribute for \texttt{<g>} into which ghost propagators are grouped
   \item[thickness] : \texttt{Number} ( length | \textbf{inherient} ) \\
    Sets the \texttt{stroke-width} attribute for \texttt{<g>} into which ghost propagators are grouped
   \item[tension] : \texttt{Number} ( parameter | \textbf{inherient} ) \\
    Sets the parameter of arc radius for ghost propagators
   \item[ratio] : \texttt{Number} ( parameter | \textbf{inherient} ) \\
    Sets the parameter of elliptical arcs for ghost propagators
   \item[arrow] : \texttt{Boolean} ( \textbf{true} | false ) \\
    Determines whether or not to show arrows for ghost propagators
   \item[clockwise] : \texttt{Boolean} ( true | \textbf{false} ) \\
    Sets the direction of arrows for arc and loop ghost propagators
   \item[dotsep] : \texttt{Number} ( length | \textbf{8} ) \\
    Sets the \texttt{stroke-dasharray} attribute for \texttt{<g>} into which ghost propagators are grouped
   \item[offset] : \texttt{Number} ( length | \textbf{5} ) \\
    Sets the \texttt{stroke-offset} attribute for \texttt{<g>} into which ghost propagators are grouped
   \item[line] : \texttt{String} ( connections ) \\
    Sets the directed edges between graph nodes for ghost lines
   \item[arc] : \texttt{String} ( connections ) \\
    Sets the directed edges between graph nodes for ghost arcs
   \item[loop] : \texttt{String} ( connections ) \\
    Sets the directed edges between graph nodes for ghost loops
  \end{description}
 \item[gluon] : \texttt{Object}
  \begin{description}
   \item[color] : \texttt{String} ( paint | \textbf{inherient} ) \\
    Sets the \texttt{stroke} attribute for \texttt{<g>} into which gluon propagators are grouped
   \item[thickness] : \texttt{Number} ( length | \textbf{inherient} ) \\
    Sets the \texttt{stroke-width} attribute for \texttt{<g>} into which gluon propagators are grouped
   \item[tension] : \texttt{Number} ( parameter | \textbf{inherient} ) \\
    Sets the parameter of arc radius for gluon propagators
   \item[clockwise] : \texttt{Boolean} ( true | \textbf{false} ) \\
    Determines whether the first wiggle starts up or down for gluon propagators
   \item[width] : \texttt{Number} ( length | \textbf{15} ) \\
    Sets the coil width of gluon propagators
   \item[height] : \texttt{Number} ( length | \textbf{15} ) \\
    Sets the coil height of gluon propagators
   \item[factor] : \texttt{Number} ( parameter | \textbf{0.75} ) \\
    Sets the factor parameter for gluon propagators
   \item[percent] : \texttt{Number} ( parameter | \textbf{0.6} ) \\
    Sets the percent parameter for gluon propagators
   \item[scale] : \texttt{Number} ( parameter | \textbf{1.15} ) \\
    Sets the scale parameter for gluon arcs and loops
   \item[line] : \texttt{String} ( connections ) \\
    Sets the directed edges between graph nodes for gluon lines
   \item[arc] : \texttt{String} ( connections ) \\
    Sets the directed edges between graph nodes for gluon arcs
   \item[loop] : \texttt{String} ( connections ) \\
    Sets the directed edges between graph nodes for gluon loops
  \end{description}
 \item[symbol] : \texttt{Object}
  \begin{description}
   \item[color] : \texttt{String} ( paint | \textbf{inherient} ) \\
    Sets the \texttt{stroke} attribute for \texttt{<g>} into which symbols are grouped
   \item[thickness] : \texttt{Number} ( length | \textbf{inherient} ) \\
    Sets the \texttt{stroke-width} attribute for \texttt{<g>} into which symbols are grouped
   \item[s1, s2, s3, \ldots] : \texttt{Array}
    \begin{description}
     \item[{sn[0]}] : \texttt{String} ( position ) \\
      Sets the coordinates of graph nodes for symbol
     \item[{sn[1]}] : \texttt{Number} ( angle ) \\
      Sets the x-axis-rotation angle for symbol
     \item[{sn[2]}] : \texttt{String} ( "arrow" | "blob" | "bubble" | "condensate" | "hadron" | "zigzag" ) \\
      Sets the symbol type
     \item[{sn[3]}] : \texttt{Number} ( parameter | \textbf{20} ) \\
      Sets the distance parameter for the symbol
     \item[{sn[4]}] : \texttt{Number} ( parameter | \textbf{4} ) \\
      Sets the height parameter for the symbol
     \item[{sn[5]}] : \texttt{Boolean} ( true | \textbf{false} ) \\
      Enables or disables a variant for the symbol
    \end{description}
  \end{description}
 \item[node] : \texttt{Object}
  \begin{description}
   \item[color] : \texttt{String} ( paint | \textbf{inherient} ) \\
    Sets the \texttt{stroke} attribute for \texttt{<g>} into which nodes are grouped
   \item[thickness] : \texttt{Number} ( length | \textbf{inherient} ) \\
    Sets the \texttt{stroke-width} attribute for \texttt{<g>} into which nodes are grouped
   \item[show] : \texttt{Boolean | String} ( \textbf{false} | "i" | "o" | "v" | "a" | ... | "iova" ) \\
    Determines whether or not to show nodes
   \item[type] : \texttt{String} ( "box" | "boxtimes" | "cross" | \textbf{"dot"} | "otimes" ) \\
    Sets the node type
   \item[radius] : \texttt{Number} ( length | \textbf{3} ) \\
    Sets the radius parameter of nodes
   \item[fill] : \texttt{String} ( paint | \textbf{"white"} ) \\
    Sets the \texttt{fill} attribute for \texttt{<g>} into which nodes are grouped
  \end{description}
 \item[label] : \texttt{Object}
  \begin{description}
   \item[color] : \texttt{String} ( paint | \textbf{inherient} ) \\
    Sets the \texttt{stroke} attribute for \texttt{<g>} into which labels are grouped
   \item[thickness] : \texttt{Number} ( length | \textbf{0} ) \\
    Sets the \texttt{stroke-width} attribute for \texttt{<g>} into which labels are grouped
   \item[fill] : \texttt{String} ( paint | \textbf{"white"} ) \\
    Sets the \texttt{fill} attribute for \texttt{<g>} into which labels are grouped
   \item[family] : \texttt{String} ( family-name | \textbf{"Georgia"} ) \\
    Sets the \texttt{font-family} attribute for \texttt{<g>} into which labels are grouped
   \item[size] : \texttt{Number} ( length | \textbf{15} ) \\
    Sets the \texttt{font-size} attribute for \texttt{<g>} into which labels are grouped
   \item[weight] : \texttt{String} ( "normal" | "bold" | "bolder" | "lighter" ) \\
    Sets the \texttt{font-weight} attribute for \texttt{<g>} into which labels are grouped
   \item[face] : \texttt{String} ( "normal" | \textbf{"italic"} | "oblique" ) \\
    Sets the \texttt{font-style} attribute for \texttt{<g>} into which labels are grouped
   \item[align] : \texttt{String} ( "start" | \textbf{"middle"} | "end" ) \\
    Sets the \texttt{text-anchor} attribute for \texttt{<g>} into which labels are grouped
   \item[t1, t2, t3, \ldots] : \texttt{Array}
    \begin{description}
     \item[{tn[0]}] : \texttt{String} ( position ) \\
      Sets the coordinates of graph nodes for label
     \item[{tn[1]}] : \texttt{String} ( text ) \\
      Sets the text of label as the content of \texttt{<tspan>}
     \item[{tn[2]}] : \texttt{Number} ( length | \textbf{18} ) \\
      Sets the \texttt{width} attribute for \texttt{<foreignObject>}
     \item[{tn[3]}] : \texttt{Number} ( length | \textbf{30} ) \\
      Sets the \texttt{height} attribute for \texttt{<foreignObject>}
    \end{description}
  \end{description}
 \item[image] : \texttt{Object}
  \begin{description}
   \item[m1, m2, m3, \ldots] : \texttt{Array}
    \begin{description}
     \item[{mn[0]}] : \texttt{String} ( position ) \\
      Sets the coordinates of position for including external image
     \item[{mn[1]}] : \texttt{String} ( file ) \\
      Sets the path for external image file
     \item[{mn[2]}] : \texttt{Number} ( length | \textbf{32} ) \\
      Sets the \texttt{width} attribute for image
     \item[{mn[3]}] : \texttt{Number} ( length | \textbf{32} ) \\
      Sets the \texttt{height} attribute for image
    \end{description}
  \end{description}
 \item[mathjax] : \texttt{Boolean} ( true | \textbf{false} ) \\
  Determines whether or not to use MathJax to typeset mathematics in labels
 \item[ajax] : \texttt{Boolean} ( true | \textbf{false} ) \\
  Determines whether or not to merge the code of external SVG image directly
\end{description}

%% file: bezier.tex
\section{Cubic B\'{e}zier Splines\label{sec:bezier}}

SVG can produce graceful curves by graphing quadratic and cubic equations.
As mentioned before, in jQuery.Feyn we use cubic B\'{e}zier segments to
approximate the photon and gluon propagators (for the line type, gluon
propagators are drawn as elliptical arc paths supported by SVG directly).
Now, we come to answer the problem how to approximate a sine curve and
an ellipse by cubic B\'{e}zier splines. By using the symmetry,
we only need to consider one-fourth of its period.

A cubic B\'{e}zier spline needs four control points $P_0$, $P_1$, $P_2$ and $P_3$.
It is obvious that the endpoints and their tangents should be accurate.
Then, we need another constraint to determine all the control points.
For approximating the sine curve, we require the curvature of the B\'{e}zier
spline in the endpoints to be accurate as well. According to the derivation in
\url{http://mathb.in/1447}, we can obtain the following control points
\[
 P_0=(0,0),\quad P_1=(\lambda p/\pi, \lambda a/2),\quad
 P_2=(2p/\pi,a),\quad P_3=(p,a),\quad \lambda=0.51128733
\]
where $p$ is one-fourth of the period of the sine curve and $a$ is the amplitude.
For approximating an ellipse with semi-major axis $a$ and semi-minor axis $b$,
we can approximate a cirlce first and then make scaling transformations.
The control points for the ellipse in the first quarter are given by
\[
 P_0=(0,b),\quad P_1=(\kappa a, b),\quad
 P_2=(a,\kappa b),\quad P_3=(a,0),\quad \kappa=0.55191502
\]
Here we have used the result in Spencer Mortensen's article:
\begin{center}
 \url{http://spencermortensen.com/articles/bezier-circle/}
\end{center}
where the constant $\kappa$ is determined from the constraint that
the maximum radial distance from the circle to the B\'{e}zier curve
must be as small as possible.

%% file: references.tex

\phantomsection